\documentclass[preprint,pteplogo]{ptephy}

\preprintnumber{} 

\usepackage{boites} 
\usepackage[T1]{fontenc}
\usepackage{lmodern}
\usepackage{textcomp}
\usepackage{braket}
\usepackage{comment}

\usepackage{bm}

\newcommand{\Eq}[1]{Eq.~(\ref{#1})}
\newcommand{\Appendix}[1]{Appendix.~\ref{#1}}
\newcommand{\Sec}[1]{Sec.~\ref{#1}}
\bibpunct{[}{]}{,}{n}{}{;}

\begin{document}

\title{Current matrix element in HAL QCD's wave function equivalent potential method}

\author{\name{Kai Watanabe}{1}, and \name{Noriyoshi Ishii}{1}\thanks{This author contributed equally to this work}}

\address{
\affil{1}{Research Center for Nuclear Physics, Osaka University,\\
10-1 Mihoga-oka, Ibaraki-shi, Osaka 567-0047, Japan}
\email{kaiw@rcnp.osaka-u.ac.jp}
}

\begin{abstract}
We give a formula to calculate a matrix element of a conserved current
in  the  effective quantum  mechanics  defined  by the  wave  function
equivalent potentials proposed by HAL QCD collaboration.
As  a first  step, a  non-relativistic field  theory with  two channel
coupling  is considered  as the  original  theory, with  which a  wave
function equivalent HAL QCD potential is obtained in a closed analytic
form.
The external field  method is used to derive the  formula by demanding
that the result  should agree with the original theory.
%
With this formula, the matrix  element is obtained by sandwiching the
effective  current  operator between  the  left  and the  right  eigen
functions of  the effective  Hamiltonian associated  with the  HAL QCD
potential.
In  addition to  the  naive one-body  current,  the effective  current
operator  contains  an  additional  two-body term  emerging  from  the
degrees of  freedom which  has been  integrated out.
\end{abstract}

\subjectindex{
  B36,   
  B64,   
  B69,   
  D00,   
  D03,   
  D32,   
  D34    
}

\maketitle


\section{Introduction}
The nuclear force plays a key role in understanding
various properties of atomic nuclei.
It is important not only for nuclear physics but also for astrophysics
such as explosion of supernova and the structure of neutron stars.
Enormous efforts  have been devoted  to investigations of  the nuclear
force,  which  includes phenomenological  studies  such  as the  meson
exchange  interactions\cite{Machleidt:1989tm},  the  chiral  effective
field  theory \cite{Epelbaum:2008ga}  and the  recently developed  QCD
based studies \cite{Ishii:2006ec}.
Since 90's,  high-precision phase  equivalent NN potentials  have been
available   \cite{Machleidt:2000ge,   Stoks:1994wp,   Wiringa:1994wb}.
These high-precision potentials are so determined as to reproduce wide
range of  nucleon-nucleon scattering  data together with  the deuteron
properties.

A lattice  QCD (LQCD) method to  determine the nuclear force  has been
recently   developed  by   HAL  QCD   collaboration\cite{Ishii:2006ec,
  Aoki:2009ji, Aoki:2011gt, HALQCD:2012aa}, which  we will refer to as
HAL QCD method.
It has been applied to many systems \cite{Aoki:2012tk, Gongyo:2017fjb}.
With   this   method,   LQCD    is   used   to   generate   equal-time
Nambu-Bethe-Salpeter (NBS) wave functions for  NN system in the center
of mass frame.
By regarding these NBS wave functions as the NN wave functions, the NN
potentials  are defined  by demanding  that these  NBS wave  functions
should  be reproduced  by Schr\"odinger  equation below  the inelastic
threshold.
We will  refer to the potential  thus defined as HAL  QCD potential or
wave-function equivalent potential.
Note that, by using LSZ reduction  formula, it is shown that these NBS
wave functions  have the asymptotic  long distance behavior,  which is
parameterized by  the scattering phase  shift $\delta$ in  exactly the
same   way  as   that  of   the  non-relativistic   quantum  mechanics
\cite{Aoki:2009ji, Lin:2001ek, Aoki:2005uf}.  For instance, for s-wave
\begin{equation}
  \left\langle 0 \left|
  N(\bm x)
  N(\bm y)
  \right|N(\bm p)N(-\bm p),in\right\rangle
  \sim
  Z e^{i\delta(\bm p)}
  \frac{
    \sin(|\bm p||\bm x - \bm y| + \delta(\bm p))
  }{
    |\bm p||\bm x - \bm y|
  }.
  \label{eq:asymptotic_form}
\end{equation}
This implies  that HAL QCD  potentials reproduce the  scattering phase
shifts together with the NBS wave functions.
It  is   therefore  a  phase   equivalent  potential  as  well   as  a
wave-function equivalent potential.

These phase equivalent  potentials can be used to  obtain an effective
NN quantum mechanics.
However,  although the  scattering phase  shifts are  guaranteed to  be
reproduced  by  these  effective  NN  quantum  mechanics,  it  is  not
straightforward to calculate matrix elements.
Note that  phase equivalent  potentials generate wave  functions whose
long distance behaviors are constrained by the scattering phase shift.
However,  there are  no constraints  imposed  on their  short and  the
medium distance behaviors.
As a  result, with the  naive formula  of matrix elements,  the result
depends on the  choice of phase equivalent  potentials, which suggests
the existence of an additional contribution to absorb the difference.
In  fact,  it  is  known  that the  electro-magnetic  current  of  the
two-nucleon system has such an additional two-body contribution, i.e.,
$J_{\mu}(\bm x) = J_{\mu}^{(1)}(\bm  x) + J_{\mu}^{(2)}(\bm x)$, where
$J_{\mu}^{(1)}(\bm  x)$ denotes  the  naive  one-body nucleon  current
while $J_{\mu}^{(2)}(\bm  x)$ denotes the additional  two-body current
which is referred to as the exchange current.
The  exchange current  is  known  to emerge  from  the charged  mesons
exchanged between the two nucleons.
Because  these  charged mesons  are  ``frozen''  in the  instantaneous
potentials, their  contribution must appear as  an additional two-body
operator in the effective NN quantum mechanics.
It is known that the exchange current gives a dominant contribution in
$d\gamma \to np$ reaction \cite{Riska:1985}.

The current conservation imposes a  constraint on the exchange current
$J^{(2)}_{\mu}(\bm x)$ as
\begin{equation}
  \bm\nabla \cdot \bm J^{(2)}(\bm x)
  =
  -i\left[
    V, \sum_{i=1,2}e_i\delta^3(\bm x - \bm r_i)
    \right],
  \label{eq:siegert_constraint}
\end{equation}
where $J_0(\bm x)  \simeq \sum_{i=1,2} e_i \delta^3(\bm x  - \bm r_i)$
is assumed with  $\bm r_i$ and $e_i \equiv  \frac{e}{2}(1 + \tau^3)_i$
being the position and the charge operator acting on the isospin space
of $i$-th nucleon, respectively \cite{Riska:1989bh, Ohta:1989rj}.
In particular, this  constraint implies
(1) $\bm  J^{(2)}(\bm x)$  does not  vanish, if  the potential  $V$ is
either isospin  dependent or  is non-local.  (2)  An explicit  form of
$\bm J^{(2)}(\bm  x)$ depends  on particular  choice of  the potential
$V$.
Although it is rather a strong  constraint, it is not strong enough to
determine the complete form of $J^{(2)}_{\mu}(\bm x)$.
In  fact, for  the  one pion  exchange potential  (OPEP)  $V =  V_{\rm
  OPEP}$,  the constraint  \Eq{eq:siegert_constraint} is  satisfied by
two  different  currents  $\bm   J^{(2)}(\bm  x)$  (i)  Sachs  current
\cite{Sachs:1948zz}  and   (ii)  one  pion  exchange   current  (OPEC)
\cite{Thompson:1973ye}.
OPEC is considered  to be reasonable, because it is  obtained by going
back to the original theory (relativistic pion-nucleon coupling model)
with  the demand  that the  Bremsstrahlung amplitude  of the  original
theory should be reproduced in the effective quantum mechanics.

The same strategy does not work for the phenomenologically constructed
phase equivalent potentials, because  their connection to the original
theory (QCD) is  unclear. (For those potentials whose  relation to QCD
is unknown,  a prescription to  introduce a current operator  which is
conserved is proposed \cite{Riska:1989bh, Ohta:1989rj}.)  In contrast,
since HAL QCD  potentials are constructed in LQCD, it  may be possible
to derive  an explicit form of  the exchange current operator  for the
HAL QCD method.
Note that once such formula is  established for the HAL QCD method, it
enables us to consider QCD matrix elements by the nuclear physics with
the  conventional   nucleon  degrees  of  freedom.    There  are  many
applications. In addition to the standard calculation of form factors,
it  can be  applied to  the  nuclear EDM  for for  physics beyond  the
standard  model \cite{Engel:2013lsa},  $np\to d\gamma$  in a  big bang
nucleosynthesis \cite{Beane:2015yha}, etc.

In this paper,  we consider a method to calculate  a matrix element in
an effective quantum mechanics assoicated with HAL QCD potentials.
As a  first step,  we restrict  ourselves to the  matrix element  of a
conserved  current.
Instead of Lorentz covariant QCD, we employ a non-relativistic Galilei
covariant      (field      theoretical)     coupled-channel      model
\cite{Sugiura:2017vwo, Birse:2012ph} as an  original theory to present
a formula to calculate the matrix element of a conserved current.
This non-relativistic model  enables us to obtain a  HAL QCD potential
in a closed analytic form, which  is used to define a non-relativistic
effective quantum mechanics.
We use  a non-relativistic original  theory, because, at  this initial
stage,  we  prefer  to  have  the formula  which  involves  as  little
approximation as possible.
%
To  obtain the  formula  to calculate  a matrix  element,  we use  the
external   field  method,   which   was  conveniently   used  in   the
Bethe-Salpeter framework \cite{Kvinikhidze:1998xn, Kvinikhidze:1999xp,
  Ishii:2000zy}.
The external  field method  enables us  to obtain  the formula  in the
effective quantum mechanics  which agrees with the  calculation in the
original theory.

The paper  is organized as follows.
In  \Sec{sec:original_theory},  a  second  quantized  non-relativistic
coupled-channel model  is introduced, which mimics  np-np$^*$ coupling
system.
In  \Sec{sec:halqcd},  the  non-relativistic  model  is  used  as  the
original theory to obtain an  effective quantum mechanics of np system
below  the  np$^*$ threshold  by  integrating  out the  closed  np$^*$
channel.
This is  done by using  HAL QCD method. We  will give a  wave function
equivalent  HAL  QCD  potential  in   a  closed  analytic  form  which
reproduces the equal-time Nambu-Bethe-Salpeter (NBS) wave functions of
open np channel.
In \Sec{sec:external_field}, we introduce  an external gauge field and
extend the HAL QCD potential in the external field.
In \Sec{sec:formula},  the external field  method is used to  derive a
formula  to calculate  the  current matrix  element  in the  effective
quantum mechanics so that the result agrees with the original theory.
Readers  may wonder  why we  employ  the np-np$^*$  coupling model  to
consider  the ``exchange  current'' instead  of those  models where  a
potential is obtained by integrating  out the exchanged mesons between
two nucleons.
This is because we determine to stick to the non-relativistic original
theory to obtain  an analytic expression of the formula  of the matrix
element.
Note that,  even with np-np$^*$  coupling model, the  two-body current
will appear after integrating out  the closed np$^*$ channel to obtain
the effective np potential.

\section{The original theory\label{sec:original_theory}} 
\subsection{Hamiltonian}
We consider a second quantized non-relativistic Hamiltonian
\begin{eqnarray}
  \hat H
  &\equiv&
  \hat T + \hat V
  \label{eq:hamiltonian}
\end{eqnarray}
where $\hat  T \equiv  \hat T_1  + \hat  T_2 +  \hat T_3$  denotes the
kinetic term with
\begin{eqnarray}
  \hat T_0
  &\equiv&
  \int d^3 x\,
  \hat\phi^\dagger_0({\bm x})
  \left(-\frac{\bm \partial^2}{2m}\right)
  \hat\phi_0(\bm x)
  \\\nonumber
  \hat T_1
  &\equiv&
  \int d^3 x\,
  \hat\phi^\dagger_1(\bm x)
  \left(-\frac{\bm\partial^2}{2m}\right)
  \hat\phi_1(\bm x)
  \\\nonumber
  \hat T_2
  &\equiv&
  \int d^3 x\,
  \hat\phi^\dagger_2(\bm x)
  \left(-\frac{\bm\partial^2}{2m} + \Delta\right)
  \hat\phi_2(\bm x).
\end{eqnarray}
$\hat V  \equiv \sum_{\alpha,\beta=1,2} \hat  V_{\alpha\beta}$ denotes
the interaction term with
\begin{eqnarray}
  \hat V_{\alpha\beta}
  &\equiv&
  \int d^3x\, d^3 y\,
  \hat \phi_0^\dagger(\bm x)
  \hat \phi_\alpha^\dagger(\bm y)
  V_{\alpha\beta}(\bm x - \bm y)
  \hat \phi_\beta(\bm y)
  \hat \phi_0(\bm x),
\end{eqnarray}
which   is    used   to   mimic   the    np-np$^*$   coupling   system
\cite{Birse:2012ph,   Sugiura:2017vwo},  where   $\hat\phi_0(\bm  x)$,
$\hat\phi_1(\bm x)$ and $\hat\phi_2(\bm  x)$ correspond to the neutron
(n),  the proton  (p) and  an excited  proton (p$^*$)  with excitation
energy $\Delta$, respectively.
(We use  a bold font for  three dimensional vectors.  A  variable with
hat ``$\hat{*}$''  is used  to indicate  that it  is a  field operator
acting on the Fock space.)
We  employ the  same non-relativistic  mass  $m$ for  all these  three
fields for Galilei covariance.
Since we do not stick to  the detail of the np-np$^*$ coupling system,
we consider the  scalar boson fields for simplicity  which satisfy the
equal-time commutation relation
\begin{equation}
  \left[\hat\phi_\alpha(\bm x), \hat\phi_\beta^\dagger(\bm y)\right]
  =
  \delta_{\alpha\beta}\delta^3(x - y).
\end{equation}
All the other combinations vanish.  

We consider eigenvalues  and eigenvectors of $\hat H$ as
\begin{equation}
  \hat H
  |n,\bm P\rangle
  =
  E_n(\bm P^2)
  |n,\bm P\rangle,
\end{equation}
where  $|n,\bm  P\rangle$  denotes  an  energy  eigenstates  with  the
normalization  $\langle m,\bm  Q|  n,\bm  P\rangle =  \delta_{mn}\cdot
(2\pi)^3\delta^3(\bm Q  - \bm P)$.
$\bm  P$ denotes  the  total  spatial momentum,  while  $n$ labels  an
``intrinsic  excitation'' in  the center  of mass  frame.  Due  to the
Galilei covariance, the energy eigenvalue $E_n(\bm P^2)$ decomposes as
\begin{equation}
  E_n(\bm P^2)
  =
  \widetilde E_n
  +
  \frac1{4m}\bm P^2,
\end{equation}
where  $\widetilde   E_n$  denotes  the  energy   of  the  ``intrinsic
excitation'' in the center of mass frame.
We will  refer to $\widetilde E_n$  as ``reduced'' energy.  (We  use a
variable with  tilde ``$\widetilde *$'' for  ``reduced'' objects which
have something to do with the center of mass frame such as the reduced
NBS wave functions  $\widetilde \psi(\bm r)$ which  will be introduced
later.)

\subsection{Conserved Currents and Conserved Charges}
The Hamiltonian $\hat H$ has  two $U(1)$ symmetries $U_{\rm n}(1)$ and
$U_{\rm p}(1)$. They are generated by the conserved charges
\begin{eqnarray}
  \hat Q_{\rm n}
  &\equiv&
  \int d^3 x\,
  \hat\phi_0^\dagger(\bm x)
  \hat\phi_0(\bm x)
  \\\nonumber
  \hat Q_{\rm p}
  &\equiv&
  \int d^3 x\,
  \left(
  \hat\phi_1^\dagger(\bm x)
  \hat\phi_1(\bm x)
  +
  \hat\phi_2^\dagger(\bm x)
  \hat\phi_2(\bm x)
  \right),
\end{eqnarray}
respectively.   $\hat Q_{\rm  n}$ corresponds  to the  conservation of
n-number, whereas $\hat Q_{\rm p}$  corresponds to the conservation of
p-number which also  counts p$^*$.  These charges  are associated with
the non-relativistic conserved currents
\begin{eqnarray}
  \hat j_{\rm n}^\mu(x)
  &\equiv&
  \hat j_0^\mu(x)
  \\
  \hat j_{\rm p}^\mu(x)
  &\equiv&
  \hat j_1^\mu(x) + \hat j_2^\mu(x),
  \label{eq:conserved_up1_current}
\end{eqnarray}
respectively with
\begin{eqnarray}
  \hat j_\alpha^0(x)
  &\equiv&
  \hat\phi_\alpha^\dagger(x)
  \hat\phi_\alpha(x)
  \\\nonumber
  \hat j_\alpha^i(x)
  &\equiv&
  \frac1{2mi}\left\{
  \hat\phi_\alpha^\dagger(x)
  \left(\partial^i \hat\phi_\alpha(x)\right)
  -
  \left(\partial^i \hat\phi_\alpha^\dagger(x)\right)
  \hat\phi_\alpha(x)
  \right\},
\end{eqnarray}
where    $\hat    \phi_{\alpha}(x)     \equiv    e^{i\hat    H    x_0}
\hat\phi_{\alpha}(\bm  x) e^{-i\hat  H x_0}$  denotes the  Heisenberg
operators for $\alpha = 0, 1, 2$.
By using  the Heisenberg's equation $i\partial_0  \hat\phi_\alpha(x) =
\left[\hat\phi_\alpha(x), \hat H\right]$, it is straightforward to see
that these  two currents  $\hat j_{\rm  n}^{\mu}(x)$ and  $\hat j_{\rm
  p}^{\mu}(x)$ conserve.

\subsection{The two-particle subspaces}
We  restrict  ourselves to  the  two-particle  subspace with  $(Q_{\rm
  n},Q_{\rm p})=(1,1)$,
which we will refer to as $\mathcal F$.
The subspace $\mathcal  F$ is spanned by all the  state vectors of the
form
\begin{equation}
  \left|\left. \psi \right\rangle\right.
  \equiv
  \int d^3 x\, d^3 y\,
  \left(
  \phi_0^\dagger(\bm x)\phi_1^\dagger(\bm y)
  \left|\left. 0 \right\rangle\right.
  \psi_1(\bm x,\bm y)
  +
  \phi_0^\dagger(\bm x)\phi_2^\dagger(\bm y)
  \left|\left. 0 \right\rangle\right.
  \psi_2(\bm x,\bm y)
  \right),
  \label{eq:two_particle_subspace_f}
\end{equation}
where $|0\rangle$  denotes the non-relativistic vacuum  defined by the
relation
\begin{equation}
  \hat\phi_\alpha(\bm x)|0\rangle  = 0,
\end{equation}
for $\alpha = 0,1,2$ and all  $\bm x \in \mathbb{R}^3$ with $\langle 0
| 0 \rangle = 1$.

We introduce a cutoff by using a projection operator
\begin{equation}
  \hat{\mathbb P}_\Lambda
  \equiv
  |0\rangle
  \langle 0|
  +
  \sum_{n}^{\widetilde E_n < \Lambda}
  \int \frac{d^3P}{(2\pi)^3}
  |n,\bm P\rangle
  \langle n, \bm P|.
\end{equation}
Note that this cutoff is Galilei covariant.
We use it to define a truncated subspace ${\mathcal F}_\Delta$ by
\begin{equation}
  {\mathcal F}_{\Delta}
  \equiv
  \hat{\mathbb{P}}_\Delta
  \cdot
  {\mathcal F}.
\end{equation}
The truncated subspace ${\mathcal F}_\Delta$ consists of all states in
$\mathcal F$ which exist below the np$^*$ threshold.
From  \Sec{sec:halqcd}, we  will use  HAL QCD  method to  construct an
effective  quantum mechanics  for states  in ${\mathcal  F}_\Delta$ by
integrating out all the states $|n, \bm P\rangle$ with $\widetilde E_n
> \Delta$.

\subsection{The Nambu-Bethe-Salpeter (NBS) wave function}
We define (equal-time) ``Nambu-Bethe-Salpeter'' (NBS) wave function as
\begin{equation}
  \psi_{\alpha}(\bm x,\bm y; t| n, \bm P)
  \equiv
  \left\langle
  0
  \left|
  \hat\phi_0(\bm x,t)
  \hat\phi_\alpha(\bm y,t)
  \right|
  n, \bm P
  \right\rangle,
\end{equation}
for $\alpha = 1,2$ and $|n,  \bm P\rangle \in \mathcal{F}$.
(If $t=0$, we  simply omit to write $t$,  i.e., $\psi_\alpha(\bm x,\bm
y|n, \bm P)  \equiv \langle 0 |  \hat\phi_0(\bm x) \hat\phi_\alpha(\bm
y) | n, \bm P \rangle$.)
NBS wave functions satisfy the coupled channel Schr\"odinger equation
\begin{eqnarray}
  \lefteqn{
    \left(
    E_n(\bm P^2)
    + \frac{\bm\partial_{\bm x}^2}{2m}
    + \frac{\bm\partial_{\bm y}^2}{2m}
    \right)
    \psi_{1}(\bm x,\bm y| n,\bm P)
  }
  \label{eq:coupled_channel}
  \\\nonumber
  &&=
  V_{11}(\bm x - \bm y)
  \psi_{1}(\bm x,\bm y| n,\bm P)
  +
  V_{12}(\bm x - \bm y)
  \psi_{2}(\bm x,\bm y| n,\bm P)
  \\[2ex]\nonumber
  \lefteqn{
    \left(
    E_n(\bm P^2)
    +\frac{\bm\partial_{\bm x}^2}{2m}
    +\frac{\bm\partial_{\bm y}^2}{2m}
    -\Delta
    \right)
    \psi_{2}(\bm x,\bm y| n,\bm P)
  }
  \\\nonumber
  &&=
  V_{21}(\bm x - \bm y)
  \psi_{1}(\bm x,\bm y| n,\bm P)
  +
  V_{22}(\bm x - \bm y)
  \psi_{2}(\bm x,\bm y| n,\bm P),
\end{eqnarray}
which  can be  verified  by sandwiching  $\left[ \hat\phi_0(\bm  x,t)
  \hat\phi_\alpha(\bm y,t),  \hat H \right]$ between  $\langle 0|$ and
$|n,\bm P\rangle$.

Due to the Galilei covariance, NBS wave functions factorize as
\begin{equation}
  \psi_{\alpha}(\bm x,\bm y,t| n,\bm P)
  =
  \widetilde \psi_{\alpha}(\bm x - \bm y, t| n)
  \exp\left( i \bm P\cdot\frac{\bm x + \bm y}{2} \right)
  \exp\left(
  -i\frac{1}{4m}\bm P^2 t
  \right),
  \label{eq:full-NBS-wave}
\end{equation}
where  $\widetilde  \psi_{\alpha}(\bm  r,t|n)$ denotes  the  NBS  wave
function in the center of mass frame
\begin{eqnarray}
  \widetilde \psi_{\alpha}(\bm r, t| n)
  &\equiv&
  \left\langle
  0
  \left|
  \hat\phi_0(\bm r/2, t)
  \hat\phi_\alpha(-\bm r/2, t)
  \right|
  n, \bm P = \bm 0
  \right\rangle,
  \label{eq:reduced_nbs}
\end{eqnarray}
which will be referred to as the reduced NBS wave function.
(Again,   if   $t=0$,   we   simply   omit   to   write   $t$,   i.e.,
$\widetilde\psi_\alpha(\bm r|n) \equiv  \langle 0| \hat\phi_0(\bm r/2)
\hat\phi_\alpha(-\bm r/2) | n, \bm P=\bm 0\rangle$.)
Reduced NBS  wave functions satisfy the  coupled channel Schr\"odinger
equation
\begin{eqnarray}
  \left(
  \widetilde E_n
  + \frac{\bm\partial^2}{2 \widetilde m}
  \right)
  \widetilde \psi_{1}(\bm r| n)
  &=&
  V_{11}(\bm r)   \widetilde \psi_{1}(\bm r| n)
  + V_{12}(\bm r) \widetilde \psi_{2}(\bm r| n)
  \label{eq:coupled_channel_reduced}
  \\\nonumber
  \left(
  \widetilde E_n
  + \frac{\bm\partial^2}{2 \widetilde m}
  - \Delta
  \right)
  \widetilde \psi_{2}(\bm r| n)
  &=&
  V_{21}(\bm r)   \widetilde\psi_{1}(\bm r| n)
  + V_{22}(\bm r) \widetilde\psi_{2}(\bm r| n),
\end{eqnarray}
where $\widetilde m \equiv m/2$ denotes the reduced mass.

\section{The HALQCD potential and the effective quantum mechanics\label{sec:halqcd}}
\subsection{The HAL QCD potentials}
We  use HAL  QCD  method to  obtain an  effective  np potential  (wave
function equivalent HAL QCD potential).
This is carried out in two steps.
We first  construct the (reduced) HAL  QCD potential in the  center of
mass  frame.  To  do this,  we require  the Schr\"odinger  equation to
reproduce  the reduced  NBS wave  functions  of np  channel below  the
np$^*$ threshold.
We then use Galilei covariance to generalize the reduced potential for
a general Galilei frame, with  which (full) NBS wave functions satisfy
the Schr\"odinger equation.

To obtain  the reduced  HAL QCD potential  $\widetilde{\mathcal V}(\bm
r,\bm  r')$, we  demand that,  for any  states $|n,\bm  P=0\rangle \in
{\mathcal  F}_{\Delta}$,  the reduced  NBS  wave  function of  the  np
channel $\widetilde\psi_1(\bm  r| n) \equiv \langle  0| \hat\phi_0(\bm
r/2)\phi_1(-\bm  r/2)|n,\bm  P=\bm 0\rangle$  satisfies  Schr\"odinger
equation
\begin{equation}
  \left(
  \widetilde{E}_n + \frac{1}{2\widetilde m}\bm\partial^2
  \right)
  \widetilde \psi_{1}(\bm{r}| n)
  = \int d^3r'
  \widetilde{\mathcal V}(\bm{r},\bm{r'})
  \widetilde \psi_{1}(\bm{r'}| n),
\label{HALQCD-eq-relative}
\end{equation}
where  $\widetilde  E_n$  denotes  the energy  eigenvalue  of  $|n,\bm
P=0\rangle$.
The demand is satisfied  by the following energy-independent non-local
potential  $\widetilde  {\mathcal V}(\bm  r,\bm  r')$ as
\begin{equation}
  \widetilde{\mathcal V}(\bm{r},\bm{r'})
  \equiv
  \sum^{\widetilde{E}_m<\Delta}_{m}
  \left(
    V_{11}(\bm{r})
    \widetilde \psi_{1}(\bm{r}| m)
    +
    V_{12}(\bm{r})
    \widetilde \psi_{2}(\bm{r}| m)
    \right)
  \widetilde \psi^\vee_{1}(\bm{r'}| m),
  \label{eq:non-local-pot}
\end{equation}
where $\widetilde  \psi_{2}(\bm r|  m) \equiv \langle  0|\hat \phi_0(\bm
r/2) \hat\phi_2(-\bm r/2)|m,\bm P=0\rangle$.
$\widetilde \psi_{1}^\vee(\bm  r|m)$ denotes  a dual  basis associated
with a linearly independent set of reduced NBS wave functions $\left\{
\left.  \widetilde  \psi_1(\bm r| m)  \right| |m,\bm P =  \bm 0\rangle
\in  {\mathcal  F}_\Delta \right\}$.   The  dual  basis satisfies  the
orthogonality relation
\begin{equation}
  \int d^3 r\,
  \widetilde \psi_{1}^\vee(\bm r|m)
  \widetilde \psi_{1}(\bm r|n)
  =
  \delta_{mn},
  \label{eq:orthogonality}
\end{equation}
for $m, n$ with $\widetilde E_m, \widetilde E_n < \Delta$.
An explicit form of dual basis is given, for instance, by
\begin{equation}
  \widetilde\psi_1^\vee(\bm r| n)
  \equiv
  \sum_{m}^{\widetilde E_m < \Delta}
  (\widetilde{\mathcal N}^{-1})_{nm}
  \widetilde\psi_{1}^*(\bm r| m),
  \label{eq:norm_kernel}
\end{equation}
where    $\widetilde{\mathcal   N}_{nm}    \equiv    \int   d^3    r\,
\widetilde\psi_{1}^*(\bm  r| n)  \widetilde\psi_{1}(\bm r|m)$  denotes
the norm kernel.
%

It  is  straightforward  to  see  \Eq{HALQCD-eq-relative},  i.e.,  the
Schr\"odinger equation with the  reduced HAL QCD potential $\widetilde
{\mathcal  V}(\bm r,\bm  r')$ is  satisfied  by the  reduced NBS  wave
functions.
For    this   purpose,    we    insert   \Eq{eq:non-local-pot}    into
\Eq{HALQCD-eq-relative} and perform the integration over $\bm r'$ as
\begin{eqnarray}
  \mbox{r.h.s. of \Eq{HALQCD-eq-relative}}
  &=&
  \int d^3 r'\,
  \sum^{\widetilde{E}_m<\Delta}_{m}
  \left(
    V_{11}(\bm{r})
    \widetilde \psi_{1}(\bm{r}| m)
    +
    V_{12}(\bm{r})
    \widetilde \psi_{2}(\bm{r}| m)
    \right)
  \widetilde \psi^\vee_{1}(\bm{r'}| m)
  \widetilde \psi_{1}(\bm r'| n)
  \nonumber
  \\
  &=&
  V_{11}(\bm{r})
  \widetilde \psi_{1}(\bm{r}| n)
  +
  V_{12}(\bm{r})
  \widetilde \psi_{2}(\bm{r}| n),
  \label{eq:rhs_reduced}
\end{eqnarray}
where the orthogonality relation  \Eq{eq:orthogonality} is used.  Thus
the Schr\"odinger  equation (\ref{HALQCD-eq-relative}) reduces  to the
coupled channel equation (\ref{eq:coupled_channel_reduced}).

We use Galilei covariance to  generalize the reduced HAL QCD potential
$\widetilde  {\mathcal  V}(\bm  r,\bm  r')$  to  the  (full)  HAL  QCD
potential ${\mathcal V}(\bm x,\bm y; \bm x',\bm y')$ by
\begin{equation}
  {\mathcal V}(\bm x,\bm y; \bm x',\bm y')
  \equiv
  \widetilde {\mathcal V}(\bm x - \bm y; \bm x' - \bm y')
  \delta^3\left(
  \frac{\bm x + \bm y}{2}
  -
  \frac{\bm x' + \bm y'}{2}
  \right).
  \label{eq:non-local-pot_full}
\end{equation}
Note that, for  any states $|n,\bm P\rangle  \in {\mathcal F}_\Delta$,
(full) NBS wave functions of the np channel $\psi_1(\bm x,\bm y|n, \bm
P)  \equiv  \langle  0|\hat\phi_0(\bm   x)\hat\phi_1(\bm  y)|  n,  \bm
P\rangle$ satisfy Schr\"odinger equation with this potential as
\begin{equation}
  \left(
  E_n(\bm P^2)
  +
  \frac1{2m}\bm\partial_{\bm x}^2
  +
  \frac1{2m}\bm\partial_{\bm y}^2
  \right)
  \psi_{1}(\bm x,\bm y| n, \bm P)
  =
  \int d^3 x'\,d^3 y'\,
  {\mathcal V}(\bm x,\bm y; \bm x',\bm y')
  \psi_{1}(\bm x',\bm y'| n, \bm P).
  \label{eq:schrodinger-full}
\end{equation}
To see this, we insert \Eq{eq:non-local-pot_full} into r.h.s.  and use
the   factorization    formula   \Eq{eq:full-NBS-wave}.     Then   the
Schr\"odinger  equation  reduces  to  \Eq{eq:coupled_channel} as
\begin{eqnarray}
  \nonumber
  \mbox{r.h.s. of \Eq{eq:schrodinger-full}}
  &=&
  \int d^3 r'\,
  \widetilde {\mathcal V}(\bm x - \bm y,\bm r')
  \widetilde \psi_{1}(\bm r'| n)
  \exp\left( i\bm P\cdot\frac1{2}(\bm x + \bm y)\right)
  \\
  &=&
  V_{11}(\bm x - \bm y)\psi_{1}(\bm x,\bm y| n, \bm P)
  +
  V_{12}(\bm x - \bm y)\psi_{2}(\bm x,\bm y| n, \bm P),
\end{eqnarray}
where the last line is obtained by using \Eq{eq:rhs_reduced}.

\subsection{The effective quantum mechanics}
The     HAL      QCD     potentials      \Eq{eq:non-local-pot}     and
\Eq{eq:non-local-pot_full}  are used  to define  an effective  quantum
mechanics.
For later convenience, we give a summary of the eigenvalue property of
the effective quantum mechanics.

We begin  with the  reduced system  in the center  of mass  frame. The
eigenvalue relations are given as
\begin{eqnarray}
  \widetilde{\mathcal H}
  \,
  \widetilde\chi_{n}^{\rm R}(\bm r)
  &=&
  \widetilde {\mathcal E}_n
  \widetilde\chi_{n}^{\rm R}(\bm r),
  \\\nonumber
  \widetilde\chi_{n}^{\rm L}(\bm r)
  \,
  \widetilde{\mathcal H}
  &=&
  \widetilde {\mathcal E}_n
  \widetilde\chi_{n}^{\rm L}(\bm r),
\end{eqnarray}
where  $\widetilde{\mathcal  H}$  denotes the  effective  Hamiltonian,
whose actions on the left and the right wave functions are defined as
\begin{eqnarray}
  \widetilde{\mathcal H}
  \,
  \widetilde \chi^{\rm R}_n(\bm r)
  &\equiv&
  - \frac{\bm \partial^2}{2 \widetilde m}
  \widetilde \chi^{\rm R}_n(\bm r)
  +
  \int d^3 r'\,
  \widetilde {\mathcal V}(\bm r,\bm r')
  \widetilde \chi^{\rm R}_n(\bm r').
  \\\nonumber
  \widetilde \chi^{\rm L}_n(\bm r)
  \,
  \widetilde{\mathcal H}
  &\equiv&
  - \frac{\bm \partial^2}{2 \widetilde m}
  \widetilde \chi^{\rm L}_n(\bm r)
  +
  \int d^3 r'\,
  \widetilde \chi^{\rm L}_n(\bm r')
  \widetilde {\mathcal V}(\bm r',\bm r).
\end{eqnarray}
$\widetilde\chi_n^{\rm  L}(\bm r)$  and $\widetilde\chi_n^{\rm  R}(\bm
r)$ denote the  left and the right eigen functions  associated with the
energy eigenvalue $\widetilde {\mathcal E}_n$.
They satisfy the orthogonality relations
\begin{eqnarray}
  \label{left-right-eigenvectors}
  \sum^{\infty}_{n=0}
  \widetilde{\chi}^{\rm R}_{n}(\bm r)
  \,
  \widetilde{\chi}^{\rm L}_{n}(\bm r')
  &=&
  \delta^3(\bm r - \bm r')
  \\\nonumber
  \int d^3 r
  \,
  \widetilde\chi^{\rm L}_n(\bm r)
  \widetilde\chi^{\rm R}_m(\bm r)
  &=&
  \delta_{nm}.
\end{eqnarray}
Note that, in  the upper relation, the summation is  not restricted to
$\widetilde E_n < \Delta$.
Since  $\widetilde{\mathcal  V}(\bm r,\bm  r')$  is  not Hermitian  in
general, $\widetilde\chi_n^{\rm L}(\bm r)$  is not a complex conjugate
of $\widetilde\chi_n^{\rm R}(\bm r)$.

We note that, since $\widetilde  {\mathcal V}$ is defined to reproduce
the NBS wave functions in the elastic region, we have
\begin{eqnarray}
  \widetilde\chi_n^{\rm R}(\bm x - \bm y)
  &=&
  \left\langle
  0
  \left|
  \hat\phi_0(\bm x)
  \hat\phi_1(\bm y)
  \right|
  n, \bm P = \bm 0
  \right\rangle
  \\\nonumber
  \widetilde{\mathcal E}_n
  &=&
  \widetilde E_n,
\end{eqnarray}
for the states $|n, \bm P=\bm 0\rangle \in {\mathcal F}_\Delta$.
Due to the upper relation, the  scattering phase shift of the original
theory is  reproduced by the  effective quantum mechanics  through the
right eigen function  $\widetilde\chi_n^{\rm R}(\bm r)$ for  the energy
region $\widetilde E_n < \Delta$.

We  then  use Galilei  covariance  to  generalize these  relations  to
arbitrary Galilei frames.  By defining
\begin{eqnarray}
  \chi^{\rm R}_{n,\bm P}(\bm x,\bm y)
  &\equiv&
  \widetilde \chi^{\rm R}_{n}(\bm x - \bm y)
  \cdot
  \exp\left( i \bm P\cdot \frac1{2}(\bm x + \bm y) \right)
  \\\nonumber
  \chi^{\rm L}_{n,\bm P}(\bm x,\bm y)
  &\equiv&
  \widetilde \chi^{\rm L}_{n}(\bm x - \bm y)
  \cdot
  \exp\left( - i \bm P\cdot\frac1{2}(\bm x + \bm y) \right),
\end{eqnarray}
the eigenvalue relations are given as
\begin{eqnarray}
  {\mathcal H}
  \,
  \chi^{\rm R}_{n,\bm P}(\bm x,\bm y)
  &=&
  {\mathcal E}_n(\bm P^2)
  \chi^{\rm R}_{n,\bm P}(\bm x,\bm y)
  \label{eq:eigenvalue_properties}
  \\\nonumber
  \chi^{\rm L}_{n,\bm P}(\bm x,\bm y)
  \,
  {\mathcal H}
  &=&
  {\mathcal E}_n(\bm P^2)
  \chi^{\rm L}_{n,\bm P}(\bm x,\bm y),
\end{eqnarray}
where $\mathcal H$ denotes the effective Hamiltonian, whose actions on
the left and right wave functions are defined as
\begin{eqnarray}
  &&
  \\
  \nonumber
  {\mathcal H}
  \,
  \chi^{\rm R}_{n,\bm P}(\bm x,\bm y)
  &\equiv&
  \left(
  -\frac{\bm\partial_{\bm x}^2}{2m}
  -\frac{\bm\partial_{\bm y}^2}{2m}
  \right)
  \chi^{\rm R}_{n,\bm P}(\bm x,\bm y)
  +
  \int d^3x'\,d^3y'\,
  {\mathcal V}(\bm x,\bm y; \bm x',\bm y')
  \chi^{\rm R}_{n,\bm P}(\bm x',\bm y')
  \\\nonumber
  \chi^{\rm L}_{n,\bm P}(\bm x,\bm y)
  \,
  {\mathcal H}
  &\equiv&
  \left(
  -\frac{\bm\partial_{\bm x}^2}{2m}
  -\frac{\bm\partial_{\bm y}^2}{2m}
  \right)
  \chi^{\rm L}_{n,\bm P}(\bm x,\bm y)
  +
  \int d^3x'\,d^3y'\,
  \chi^{\rm L}_{n,\bm P}(\bm x',\bm y')  
  {\mathcal V}(\bm x',\bm y'; \bm x,\bm y).
\end{eqnarray}
$\chi^{\rm L}_n(\bm  x,\bm y)$ and  $\chi^{\rm R}_n(\bm x,\bm  y)$ are
the left  and right eigen functions, respectively,  associated with the
energy eigenvalue ${\mathcal  E}_n(\bm P^2) \equiv \widetilde{\mathcal
  E}_n + \frac1{4m}\bm P^2$.
They satisfy the orthogonality relations
\begin{equation}
  \sum_{n=0}^{\infty}
  \int \frac{d^3 P}{(2\pi)^3}\,
  \chi^{\rm R}_{n,\bm P}(\bm x',\bm y')
  \chi^{\rm L}_{n,\bm P}(\bm x,\bm y)
  =
  \delta^3(\bm x' - \bm x)
  \delta^3(\bm y' - \bm y),
\end{equation}
and
\begin{equation}
  \int d^3 x\,d^3 y\,
  \chi^{\rm L}_{n',\bm P'}(\bm x,\bm y)
  \chi^{\rm R}_{n,\bm P}  (\bm x,\bm y)
  =
  \delta_{n'n}
  \cdot
  (2\pi)^3 \delta^3(\bm P'-\bm P).
\end{equation}
Needless  to say,  for  the  states $|n,  \bm  P\rangle \in  {\mathcal
  F}_{\Delta}$,  the  right eigen functions  agree  with  the NBS  wave
functions as
\begin{eqnarray}
  \chi^{\rm R}_{n,\bm P}(\bm x,\bm y)
  &=&
  \left\langle
  0
  \left|
  \hat\phi_0(\bm x)
  \hat\phi_1(\bm y)
  \right|
  n, \bm P
  \right\rangle
  \\\nonumber
  {\mathcal E}_n(\bm P^2)
  &=&
  E_n(\bm P^2).
\end{eqnarray}

\section{The external field\label{sec:external_field}}
In  order to  consider the  matrix  element of  the conserved  $U_{\rm
  p}(1)$ current in the effective quantum mechanics,
we  employ  the external  field  method.
We  introduce  an external  $U_{\rm  p}(1)$  gauge field  $A_{\mu}(\bm
x,t)$, ($\mu=0,1,2,3$) and  demand that the response  of the effective
quantum mechanics  to the  external field  should be  the same  as the
original theory.
To do this, we construct a HAL QCD potential in the external field.
To  avoid  unnecessary  complexity,  we restrict  ourselves  to  those
external field  $A_{\mu}(\bm x,  t)$ which are  non-zero only  for the
time region $t > 0$.

\subsection{Hamiltonian in the external field}
We consider the coupling of the original theory to the external field.
In order for the coupling to  be consistent with the conserved $U_{\rm
  p}(1)$  current  of \Eq{eq:conserved_up1_current},  the  Hamiltonian
should be
\begin{eqnarray}
  \hat H[A_t]
  &\equiv&
  \hat T[A_t]
  + \hat V
  \label{eq:hamiltonian}
\end{eqnarray}
where the kinetic term $\hat T \equiv  \hat T_0 + \hat T_1[A_t] + \hat
T_{2}[A_t]$ couples to the external fields as
\begin{eqnarray}
  \hat T_0
  &\equiv&
  \int d^3 x\,
  \hat\phi^\dagger_0(\bm x)
  \left(
  -\frac{\bm\partial^2}{2m}
  \right)
  \hat\phi_0(\bm x)
  \\\nonumber
  \hat T_1[A_t]
  &\equiv&
  \int d^3 x\,
  \hat\phi^\dagger_1(\bm x)
  \left(
  -\frac{(\bm\partial - i\bm A(\bm x, t))^2}{2m}
  -A_{0}(\bm x,t)
  \right)
  \hat\phi_1(\bm x)
  \\\nonumber
  \hat T_2[A_t]
  &\equiv&
  \int d^3 x\,
  \hat\phi^\dagger_2(\bm x)
  \left(
  -\frac{(\bm\partial - i\bm A(\bm x, t))^2}{2m}
  -A_0(\bm x,t)
  + \Delta
  \right)
  \hat\phi_2(\bm x),
\end{eqnarray}
whereas the  interaction term  $\hat V  \equiv \sum_{\alpha,\beta=1,2}
\hat V_{\alpha\beta}$ does not couple.  (If $\hat V$ were to couple to
the external field, the conserved $U_{\rm p}(1)$ current would need an
additional term.)
The subscript ``$t$'' of the external  field $A_t$ is used to indicate
that $\hat H[A_t]$  depends on $A(\bm x,t)$ of time-slice  $t$.
The Schr\"odinger equation is given as
\begin{equation}
  i \frac{\partial}{\partial t}
  |\psi, A; t\rangle
  =
  \hat H[A_t]
  |\psi, A; t\rangle.
\end{equation}

\subsection{Truncated Hamiltonian and Truncated time-evolution in the external field}
Time-dependence of the external field causes an unwanted transition to
np$^*$ above the inelastic threshold, which is harmful in constructing
a low-energy  effective quantum mechanics below  the np$^*$ threshold.
In  order to  suppress  such  an unwanted  transition,  we insert  the
projection  operator $\hat{\mathbb{P}}_\Delta$  at every  step of  the
time  evolution.  This  is done  by replacing  the Hamiltonian  by the
truncated Hamiltonian
\begin{equation}
  \hat H_\Delta[A_t]
  \equiv
  \hat{\mathbb{P}}_\Delta
  \hat H[A_t]
  \hat{\mathbb{P}}_\Delta.
\end{equation}
$\hat  H_\Delta[A_t]$  generates  a   time-evolution,  which  will  be
referred   to  as   the  truncated   time-evolution.   The   truncated
time-evolution  is  denoted  by  $\hat  U_\Delta(t,s;  A)$,  which  is
explicitly expressed as a time-ordered product as
\begin{equation}
  \hat U_\Delta(t,s;A)
  \equiv
  \sum^{\infty}_{n=0}
  (-i)^n
  \int^{t}_{s}dt_n
  \int^{t_n}_{s}dt_{n-1}
  \cdots
  \int^{t_2}_{s} dt_1
  \,
  \hat H_\Delta[A_{t_n}]
  \hat H_\Delta[A_{t_{n-1}}]
  \cdots
  \hat H_\Delta[A_{t_1}].
  \label{eq:projected-time-evolution}
\end{equation}
Note  that $U_{\Delta}(t,s;  A)$ is  a solution  of the  initial value
problems
\begin{eqnarray}
  i\frac{\partial}{\partial t}
  \hat U_{\Delta}(t,s; A)
  &=&
  \hat H_\Delta[A_t]
  \hat U_{\Delta}(t,s; A)
  \label{eq:projected-time-evolution-diff-eq}
  \\\nonumber
  i\frac{\partial}{\partial s}
  \hat U_{\Delta}(t,s; A)
  &=&
  - \hat U_{\Delta}(t,s; A)
  \hat H_{\Delta}[A_s],
\end{eqnarray}
with $\hat U_{\Delta}(t=s,s, A) = \hat{\mathbb{I}}$.

\subsection{The truncated NBS wave functions in the external field}
The  equal-time NBS  wave  function  in the  external  field with  the
truncated time-evolution is defined as
\begin{equation}
  \psi^{(\Delta)}_{\alpha}(\bm x,\bm y,t;A| n,\bm P)
  \equiv
  \left\langle
  0
  \left|
  \hat \phi_{0}^{(\Delta)}(\bm x,t;A)
  \hat \phi_{\alpha}^{(\Delta)}(\bm y,t;A)
  \right|
  n, \bm P
  \right\rangle,
  \label{projected-NBS-op}
\end{equation}
for $\alpha = 1, 2$,
where $\hat\phi_\alpha^{(\Delta)}(\bm x,t;  A)$ denotes the Heisenberg
operator with the truncated time-evolution as
\begin{equation}
  \hat \phi_{\alpha}^{(\Delta)}(\bm x,t;A)
  \equiv
  \hat U_\Delta(0,t;A)
  \hat\phi_\alpha(\bm x)
  \hat U_\Delta(t,0;A).
\label{projected-heisenberg-op}
\end{equation}
We  will refer  to  \Eq{projected-NBS-op} as  the  truncated NBS  wave
function.
Note that, unlike \Eq{eq:full-NBS-wave}, $\psi^{(\Delta)}_{\alpha}(\bm
x, \bm  y, t;  A| n,\bm  P)$ does  not factorize  any more,  since the
external field breaks the Galilei covariance.

As  will be  shown in  \Appendix{sec:derivation_coupled_equation}, the
truncated NBS wave functions satisfy the coupled channel Schr\"odinger
equations
\begin{eqnarray}
  \lefteqn{
    \left(
    i \partial_t
    + \frac{\bm\partial_{\bm x}^2}{2m}
    + \frac{\bm D_{\bm y}^2}{2m}
    + A_0(\bm y, t)
    \right)
    \psi^{(\Delta)}_{1}(\bm x,\bm y,t;A| n, \bm P)
  }
  \label{eq:projected-schroedinger}
  \\\nonumber
  &=&
  \int d^3x'\,d^3y'
  \\\nonumber
  &\times&
  \left\{
  V_{11}(\bm x,\bm y; \bm x',\bm y'; A_t)
  \psi^{(\Delta)}_{1}(\bm x',\bm y',t;A| n, \bm P)
  +
  V_{12}(\bm x,\bm y; \bm x',\bm y'; A_t)
  \psi^{(\Delta)}_{2}(\bm x',\bm y',t;A| n, \bm P)
  \right\}
  \\[2ex]\nonumber
  \lefteqn{
    \left(
    i \partial_t
    + \frac{\bm\partial_{\bm x}^2}{2m}
    + \frac{\bm D_{\bm y}^2}{2m}
    + A_0(\bm y, t)
    - \Delta
    \right)
    \psi^{(\Delta)}_{2}(\bm x,\bm y,t;A| n, \bm P)
  }
  \\\nonumber
  &=&
  \int d^3x'\,d^3y'
  \\\nonumber
  &\times&
  \left\{
  V_{21}(\bm x,\bm y; \bm x',\bm y'; A_t)
  \psi^{(\Delta)}_{1}(\bm x',\bm y',t;A| n,\bm P)
  + V_{22}(\bm x,\bm y; \bm x',\bm y'; A_t)
  \psi^{(\Delta)}_{2}(\bm x',\bm y',t;A| n,\bm P)
  \right\},
\end{eqnarray}
where $\bm  D_{\bm y} \equiv \bm\partial_{\bm  y} - i \bm  A(\bm y,t)$
denotes the covariant derivative,
and   $V_{\alpha\beta}(\bm  x,\bm   y;   \bm   x',\bm  y';   A_t)$
($\alpha,\beta=1,2$) is defined as
\begin{equation}
  V_{\alpha\beta}(\bm x,\bm y; \bm x',\bm y'; A_t)
  \equiv
  V_{\alpha\beta}(\bm x - \bm y)
  \delta^3(\bm x - \bm x')
  \delta^3(\bm y - \bm y')
  +
  \Delta V_{\alpha\beta}(\bm x,\bm y; \bm x',\bm y'; A_t),
\end{equation}
with
\begin{equation}
  \Delta V_{\alpha\beta}(\bm x,\bm y;\bm{x'},\bm{y'};A_t)
  \equiv
  -
  \left\langle 0 \left|
  \hat \phi_0(\bm x) 
  \hat \phi_\alpha(\bm y)
  \left(
  \hat{\mathbb{I}}
  -
  \hat{\mathbb{P}}_\Delta
  \right)
  \hat H[A_t]
  \hat{\mathbb{P}}_\Delta
  \hat \phi_0(\bm x')
  \hat \phi_\beta (\bm y')
  \right| 0\right\rangle
\end{equation}
We  give several  comments.
(1) Since $\hat \phi_0(\bm x,t)$ does not have $Q_{\rm p}$ charge, the
  derivative  for  $\bm  x$  remains  to be  an  ordinary  one,  i.e.,
  $\bm\partial_{\bm x}$ in \Eq{eq:projected-schroedinger}.
(2)  The additional  term $\Delta  V_{\alpha\beta}(\cdots)$ originates
  from the cutoff $\hat{\mathbb{P}}_{\Delta}$ in $\hat H_\Delta[A_t]$.
(3) $\Delta V_{\alpha\beta}(\cdots)$ vanishes  for $A(\bm x, t)=0$ due
  to  the factor  $(\hat{\mathbb{I}}  - \hat{\mathbb{P}}_\Delta)  \hat
  H[A_t]  \hat{\mathbb{P}}_\Delta$.
(4) $\Delta V_{\alpha\beta}(\bm x,\bm y;  \bm x',\bm y'; A_t)$ depends
  on $A_{\mu}(\bm x,t)$ of time-slice $t$.

\subsection{The HAL QCD potential in the external field}
We define HAL  QCD potential in the presence of  the external field by
demanding  that,  for  any  states $|n,  \bm  P\rangle  \in  {\mathcal
  F}_\Delta$,   the  time-dependent   Schr\"odinger  equation   should
reproduce the truncated NBS wave functions for np channel as
\begin{eqnarray}
  \lefteqn{
    \left(
    i \partial_t
    + \frac{\bm\partial_{\bm x}^2}{2m}
    + \frac{\bm D_{\bm y}^2}{2m}
    + A_0(\bm y, t)
    \right)
    \psi^{(\Delta)}_{1}(\bm x,\bm y,t;A| n,\bm P)
  }
  \label{HALQCD-eq}
  \\\nonumber
  &= &
  \int d^3x'\,d^3y'\,
  {\mathcal V}(\bm{x},\bm{y};\bm{x'},\bm{y'};A_t)
  \psi^{(\Delta)}_{1}(\bm {x'},\bm {y'} , t ; A| n,\bm P).
\end{eqnarray}
We note  that, due to  the time-dependence  of the external  field, we
need to  use the time-dependent  Schr\"odinger equation to  define HAL
QCD potential.
The demand is satisfied by
\begin{eqnarray}
  \lefteqn{
    {\mathcal V}(\bm x,\bm y; \bm x',\bm y'; A_t)
  }
  \label{eq:halqcd_potential_in_external_field}
  \\\nonumber
  &\equiv&
  \sum_{m=0}^{\widetilde E_m<\Delta}
  \int d^3 x''\,d^3y''
  \\\nonumber
  &&
  \left\{
  V_{11}(\bm x,\bm y; \bm x'',\bm y''; A_t)
  \widetilde\psi_{1}(\bm x''-\bm y''|m)
  +
  V_{12}(\bm x,\bm y; \bm x'',\bm y''; A_t)
  \widetilde\psi_{2}(\bm x''-\bm y''|m)
  \right\}
  \\\nonumber
  &&\times
  \widetilde\psi^\vee_1(\bm x'-\bm y'|m)
  \delta^3\left(
  \frac{\bm x'' + \bm y''}{2}
  - \frac{\bm x' + \bm y'}{2}
  \right),
\end{eqnarray}
where    $\widetilde    \psi_{\alpha}(\bm    x''-\bm    y''|m)$    and
$\widetilde\psi_1^\vee(\bm x' - \bm y'|m)$ denote the reduced NBS wave
function  and  the  dual  basis  defined  at  \Eq{eq:reduced_nbs}  and
\Eq{eq:norm_kernel}, respectively, in the absence of external field.
The    proof     is    straightforward,    which    is     given    in
\Appendix{sec:proof_mathcal_v}.

\section{The current matrix element in the effective quantum mechanics
\label{sec:formula}}
In    \Eq{HALQCD-eq}   in    \Sec{sec:external_field},   we    derived
Schr\"odinger equation in the external field which is satisfied by the
truncated NBS wave functions.
In  this section,  we  use  this Schr\"odinger  equation  to derive  a
formula to calculate a current matrix element in the effective quantum
mechanics associated with HAL QCD potentials.
In  \Sec{sec:formula}, we  will just  give the  formula together  with
several  remarks.  The  derivation of  the  formula will  be given  in
\Sec{sec:derivation}.
\subsection{The formula to calculate a matrix element\label{sec:formula}}
Suppose  that we  have the  potential ${\mathcal  V}(\bm x,\bm  y; \bm
x',\bm y';  A_t)$ with  which Schr\"odinger  equation in  the external
field is  satisfied by the truncated  NBS wave function of  np channel
for any states $|n,\bm P\rangle \in {\mathcal F}_{\Delta}$ in the form
\Eq{HALQCD-eq}.
Then the matrix  element of the current $\hat  j_{\rm p}^{\mu}(\bm z)$
is calculated for  any states $|m,\bm Q\rangle$,  $|n,\bm P\rangle \in
{\mathcal  F}_\Delta$ in  the effective  np quantum  mechanics by  the
formula
\begin{eqnarray}
  \lefteqn{
    \left\langle
    m, \bm Q
    \left|
    \hat j_{\rm p}^{\mu}(\bm z)
    \right|
    n, \bm P
    \right\rangle
  }
  \label{eq:formula}
  \\\nonumber
  &=&
  \int d^3x\,d^3y
  \int d^3x'\,d^3y'\,
  \chi^{\rm L}_{m,\bm Q}(\bm x,\bm y)
  K^{\mu}(\bm x,\bm y; \bm x',\bm y'; \bm z)
  \chi^{\rm R}_{n,\bm P}(\bm x',\bm y'),
\end{eqnarray}
where $\chi^{\rm  L}_{m,\bm Q}(\bm x,\bm y)$  and $\chi^{\rm R}_{n,\bm
  P}(\bm x, \bm  y)$ denote the left and the  right eigen functions of
the effective Hamiltonian $\mathcal H$  in the absence of the external
field (See \Eq{eq:eigenvalue_properties}).
$K^{\mu}(\bm x,\bm  y; \bm  x',\bm y'; \bm  z)$ denotes  the effective
current operator, which is defined by
\begin{eqnarray}
  \lefteqn{
    K^{0}(\bm x,\bm y; \bm x',\bm y'; \bm z)
    \delta(t - z_0)
  }
  \label{eq:def_of_Kmu}
  \\\nonumber
  &\equiv&
  -\delta^3(\bm z - \bm y)
  \delta^3(\bm x - \bm x')
  \delta^3(\bm y - \bm y')
  \delta(t - z_0)
  +
  \left.
  \frac{\delta {\mathcal V}(\bm x, \bm y; \bm x',\bm y'; A_t)}{\delta A_0(\bm z,z_0)}
  \right|_{A\equiv 0}
  \\\nonumber
  \lefteqn{
    K^{i}(\bm x,\bm y; \bm x',\bm y'; \bm z)
    \delta(t - z_0)
  }
  \\\nonumber
  &\equiv&
  -\frac{\overleftrightarrow{\bm \partial}_{\bm z}^i}{2mi}
  \delta^3(\bm z - \bm y)
  \delta^3(\bm x - \bm x')
  \delta^3(\bm y - \bm y')
  \delta(t - z_0)
  +
  \left.
  \frac{\delta {\mathcal V}(\bm x, \bm y; \bm x',\bm y'; A_t)}{\delta A_i(\bm z,z_0)}
  \right|_{A\equiv 0},
\end{eqnarray}
with     $\overleftrightarrow{\bm     \partial}_{\bm     z}     \equiv
\overrightarrow{\bm    \partial}_{\bm    z}    -    \overleftarrow{\bm
  \partial}_{\bm z}$.
The derivation of \Eq{eq:formula} is given in \Sec{sec:derivation}.

We give several remarks.
\begin{enumerate}
\item
  In the conventional quantum  mechanics, matrix elements are obtained
  by sandwiching  an operator  with a state  vector and  its Hermitian
  conjugate.   In   contrast,  in  our  effective   quantum  mechanics
  associated with the HAL QCD  potential, an operator is sandwiched by
  the  left and  right eigen  functions of  the effective  Hamiltonian
  $\mathcal H$.
  Since HAL QCD potentials are not Hermitian in general, this could be
  a natural generalization.

\item
  The first  terms of the  effective current operator  $K^{\mu}(\bm x,
  \bm  y; \bm  x',\bm y';  \bm z)$  correspond to  the naive  one-body
  current carried by a single proton as
  \begin{eqnarray}
    \left\langle m,\bm Q\left|
    \hat j_{\rm p}^{0}(\bm z)
    \right| n, \bm P\right\rangle_{\rm naive}
    &=&
    -
    \int d^3 x\,
    \chi^{\rm L}_{m,\bm Q}(\bm x, \bm z)
    \chi^{\rm R}_{n,\bm P}(\bm x, \bm z)
    \\\nonumber
    \left\langle m,\bm Q\left|
    \hat j_{\rm p}^{i}(\bm z)
    \right| n, \bm P\right\rangle_{\rm naive}
    &=&
    -\frac1{2mi}
    \int d^3 x\,
    \chi^{\rm L}_{m,\bm Q}(\bm x, \bm z)
    \overleftrightarrow{\bm \partial}^i_{\bm z}\,
    \chi^{\rm R}_{n,\bm P}(\bm x, \bm z)
  \end{eqnarray}
  In  contrast,  the second  terms  are  the two-body  current,  which
  correspond to ``exchange current''.  The two-body currents originate
  from  the  states  above  the   np$^*$  threshold  which  have  been
  integrated out during the construction of the HAL QCD potential.

\item
  In a realistic  situation, HAL QCD potentials is  constructed not by
  the method  which was employed in  the previous sections but  by the
  derivative  expansion  using  the   NBS  wave  functions  as  inputs
  \cite{Aoki:2009ji, Sugiura:2017vwo}.
  However, as we shall see  in \Sec{sec:derivation}, the derivation of
  the formula  \Eq{eq:formula} given in \Sec{sec:derivation}  does not
  depend  on  how   HAL  QCD  potential  in  the   external  field  is
  constructed.
  All we  need to use  the formula is that  there is a  potential with
  which Schr\"odinger equation  in the external field  is satisfied by
  the truncated NBS wave functions.

\item
  It would  be interesting to  argue the gauge covariance  property as
  was done  in Ref.\cite{Ohta:1989rj}.   However, the cutoff  which we
  have introduced in the  Hamiltonian makes the situation complicated.
  Since  the  same matrix  elements  as  the  original theory  can  be
  reproduced, we do not stick too much to this point in this paper.

\item
  Application of two functional  derivative $\delta/\delta A_{\mu}$ to
  \Eq{eq:schrodinger_in_external_field}  in \Sec{sec:derivation}  does
  not  lead  to $\langle  m,\bm  Q|  j_{\rm p}^{\mu}(\bm  z_1)  j_{\rm
    p}^{\nu}(\bm  z_2) |  n, \bm  P\rangle$ but  to $\langle  m,\bm Q|
  j_{\rm  p}^{\mu}(\bm  z_1) \mathbb{P}_{\Delta}  j_{\rm  p}^{\nu}(\bm
  z_2) | n, \bm P\rangle$.  To calculate the former matrix element, an
  additional consideration is needed.
\end{enumerate}

\subsection{The derivation \label{sec:derivation}}
For notational convenience, we arrange \Eq{HALQCD-eq} as
\begin{equation}
  \left(
  i\partial_t
  - {\mathcal H}[A_t]
  \right)
  \psi_1^{(\Delta)}(\bm x, \bm y, t; A| n, \bm P)
  =
  0,
  \label{eq:schrodinger_in_external_field}
\end{equation}
where  the  effective  Hamiltonian  ${\mathcal H}[A_t]$  acts  on  the
truncated NBS wave function as
\begin{eqnarray}
  {\mathcal H}[A_t]
  \psi^{(\Delta)}_1(\bm x, \bm y, t; A| n, \bm P)
  &\equiv&
  -
  \left(
  \frac1{2m}\bm\partial_{\bm x}^2
  + \frac1{2m}\bm D_{\bm y}^2
  + A_0(\bm y, t)
  \right)
  \psi_1^{(\Delta)}(\bm x, \bm y, t; A| n, \bm P)
  \nonumber
  \\
  &+&
  \int d^3x'\,d^3y'\,
  {\mathcal V}(\bm x,\bm y; \bm x',\bm y'; A_t)
  \psi_1^{(\Delta)}(\bm x', \bm y', t; A| n, \bm P).
\end{eqnarray}
We apply the functional  derivative $\delta/\delta A_{\mu}(\bm z,z_0)$
to both  sides of  \Eq{eq:schrodinger_in_external_field} and  then set
the external field $A_{\mu}\equiv 0$ to have
\begin{eqnarray}
  \lefteqn{
    \left.
    \left(i\partial_t - {\mathcal H}\right)
    \frac{\delta \psi^{(\Delta)}_1(\bm x,\bm y, t; A|n,\bm P)}{\delta A_{\mu}(\bm z,z_0)}
    \right|_{A\equiv 0}
  }
  \label{eq:der_psi_der_a}
  \\\nonumber
  &=&
  \int d^3x'\,d^3y'\,
  K^{\mu}(\bm x, \bm y; \bm x',\bm y'; \bm z)
  \delta(t - z_0)
  \psi_1(\bm x',\bm y', t| n,\bm P),  
\end{eqnarray}
where
\begin{eqnarray}
  \lefteqn{
    K^{\mu}(\bm x,\bm y; \bm x',\bm y'; \bm z)
    \delta(t - z_0)
  }
  \\\nonumber
  &\equiv&
  \frac{\delta}{\delta A_{\mu}(\bm z,z_0)}
  \left[
    \left(
    -\frac1{2m}
    \left(
    \bm \partial_{\bm y} - i \bm A(\bm y, t)
    \right)^2
    - A_0(\bm y,t)
    \right)
    \delta^3(x - x')
    \delta^3(y - y')
    +
    {\mathcal V}(\bm x,\bm y; \bm x',\bm y'; A_t)
    \right]_{A\equiv 0},
\end{eqnarray}
which  reduces  to  the  explicit expression  of  $K^{\mu}$  given  in
\Eq{eq:def_of_Kmu}.

In the original theory,  $\delta \psi^{(\Delta)}/\delta A_{\mu}(z)$ is
expressed as
\begin{eqnarray}
  \lefteqn{
    \left.
    \frac{\delta \psi_{1}^{(\Delta)}(\bm x,\bm y, t; A|n, \bm P)}{\delta A_{\mu}(z)}
    \right|_{A\equiv 0}
  }
  \label{eq:der_psi_der_a_lhs}
  \\\nonumber
  &=&
  i\theta(z_0)
  \theta(t-z_0)
  \left\langle
  0
  \left|
  \hat \phi_0(\bm x)
  \hat \phi_1(\bm y)
  e^{i(t - z_0)\hat H}
  \hat{\mathbb P}_{\Delta}
  j^{\mu}_{\rm p}(\bm z)
  \hat{\mathbb P}_{\Delta}
  e^{iz_0\hat H}
  \right|
  n, \bm P
  \right\rangle
  \\\nonumber
  &=&
  i\theta(z_0)
  \theta(t - z_0)
  \sum_{m}^{\tilde E_m<\Delta}
  \int \frac{d^3Q}{(2\pi)^3}
  \chi_{m,\bm Q}^{\rm R}(\bm x,\bm y)
  e^{i(t-z_0)E_n(\bm Q^2)}
  \left\langle m,\bm Q\left|
  \hat j^{\mu}_{\rm p}(\bm z)
  \right| n,\bm P\right\rangle
  e^{iz_0E_n(\bm P^2)},
\end{eqnarray}
where we used \Eq{eq:der_U_der_a} to derive the second line.
The existence of $\theta(t-z_0)$ indicates
\begin{equation}
  \lim_{t\to -\infty}
  \left.
  \frac{\delta \psi_1^{(\Delta)}(\bm x,\bm y, t; A|n, \bm P)}{\delta A_{\mu}(z)}
  \right|_{A\equiv 0}
  =
  0.
  \label{eq:initial_value}
\end{equation}

To   express   $\delta\psi_1^{(\Delta)}/\delta  A_{\mu}(z)$   in   the
effective quantum  mechanics, we solve \Eq{eq:der_psi_der_a}  with the
initial  value \Eq{eq:initial_value}  by  using  the retarded  Green's
function of $\mathcal H$ given in \Eq{eq:green_function} to have
\begin{eqnarray}
  \lefteqn{
    \left.
    \frac{
      \delta \psi_{1}^{(\Delta)}(\bm x,\bm y, t; A|n, \bm P)
    }{\delta A_{\mu}(z)}
    \right|_{A\equiv 0}
  }
  \label{eq:der_psi_der_a_rhs}
  \\\nonumber
  &=&
  \int dt''
  \int d^3x''\,d^3y''
  \int d^3x'\,d^3y'
  \\\nonumber
  &\times&
  G(\bm x,\bm y, t; \bm x'',\bm y'', t'')
  K^{\mu}(\bm x'', \bm y''; \bm x',\bm y'; \bm z)
  \delta(t'' - z_0)
  \psi_1(\bm x',\bm y', t''| n,\bm P)
  \\\nonumber
  &=&
  -i\theta(t-z_0)
  \sum_{m}^{\infty}
  \int \frac{d^3Q}{(2\pi)^3}
  \chi_{m,\bm Q}^{\rm R}(\bm x,\bm y)
  e^{-iE_m(\bm Q^2)(t - z_0)}
  \\\nonumber
  &\times&
  \int d^3x''\,d^3y''
  \int d^3x'\,d^3y'\,
  \chi_{m,\bm Q}^{\rm L}(\bm x'',\bm y'')
  K^{\mu}(\bm x'',\bm y''; \bm x',\bm y';\bm z)
  \chi_{n,\bm P}^{\rm R}(\bm x',\bm y')
  e^{-i E_n(\bm P^2)z_0}.
\end{eqnarray}
By comparing  \Eq{eq:der_psi_der_a_lhs} and \Eq{eq:der_psi_der_a_rhs},
we arrive at the formula \Eq{eq:formula}.

\section{Summary and conclusion}
We have  considered how  to deal  with a matrix  element in  HAL QCD's
potential method.
HAL QCD  method is a lattice  QCD (LQCD) method to  obtain a potential
(HAL QCD potential) which is faithful to the scattering phase shift.
This is supported by the fact that
the HAL QCD  potential is defined by  demanding Schr\"odinger equation
should  reproduce  the   equal-time  Nambu-Bethe-Salpeter  (NBS)  wave
functions,
and the NBS  wave functions contain the scattering phase  shift in its
long distance part  in exactly the same way as  that of the scattering
wave functions of the non-relativistic quantum mechanics.
Therefore  the  effective  NN  quantum mechanics  associated  HAL  QCD
potential   is   supported   to   reproduce   the   scattering   phase
shift. However, an additional consideration is needed to calculate the
matrix elements.
In fact, there have been  no arguments concerning the relation between
the matrix elements calculated from the effective NN quantum mechanics
associated with the HAL QCD potential and QCD, the original theory.

As a first step to considering a  matrix element in HAL QCD method, we
have considered a simplified  non-relativistic field theoretical model
instead of Lorentz covariant QCD.
%
%
We have  employed a two-channel  coupling model as the  original theory
where np-np$^*$  coupling is  mimicked (np-np$^*$ coupling  model).
By integrating out  the closed np$^*$ channel with HAL  QCD method, we
have obtained an  effective np potential (HAL QCD  potential) which is
used to define the effective np quantum mechanics.
Due  to  the simplicity  of  our  np-np$^*$  coupling model,  we  have
obtained the HAL QCD potential in a closed analytic form.

We  have used  the external  field method  and obtained  a formula  to
calculate a matrix element of a  conserved current in the effective np
quantum mechanics
by demanding that  the response of the effective  quantum mechanics to
the external field is the same as that of the original theory.
With our formula, the matrix element is calculated by sandwiching the
effective  current  operator between  the  left  and the  right  eigen
functions of the effective np Hamiltonian.
The  effective  current  operator  consists of  two  parts  (1)  naive
one-body current and (2) the two-body current which corresponds to the
exchange  current.  In  our  np-np$^*$ coupling  model,  the two  body
current emerges from the states  above the np$^*$ threshold which have
been integrated out when obtaining the effective np potential.

To extend the formula for QCD, several generalizations are needed.
To use the formula in  relativistic original theories, it is necessary
to generalize the HAL QCD potential for a boosted Lorentz frame.
Note that it is only in the  center of mass frame where the asymptotic
forms of  NBS wave functions \Eq{eq:asymptotic_form}  agree with those
of the scattering wave functions of the quantum mechanics.
To consider  a system  of composite  particles, we  have to  take into
account the form factors.
To use the formula in LQCD, we have to deal with the time-evolution in
an external field  with a cutoff.
To do these  things, we may have to  introduce several approximations.

\section*{Acknowledgment}

We  thank  Prof.  M.~Oka,  Prof.   W.~Bentz  and Dr.  N.~Yamanaka  for
discussions.   This  work  was  supported by  Japan  Society  for  the
Promotion of Science KAKENHI Grands No. JP25400244, and by Ministry of
Education, Culture, Sports, Science and Technology as ``Priority Issue
on  Post-K  computer''  (Elucidation   of  the  Fundamental  Laws  and
Evolution  of  the Universe)  and  Joint  Institute for  Computational
Fundamental Science.


%

\appendix
\section{Derivation of \Eq{eq:projected-schroedinger}}
\label{sec:derivation_coupled_equation}
We prove  that the  truncated NBS wave  functions satisfy  the coupled
channel Schr\"odinger equations (\ref{eq:projected-schroedinger}).
By using \Eq{eq:projected-time-evolution-diff-eq}, the time derivative
of  $\hat\phi_0^{(\Delta)}(\bm x,t;  A) \hat\phi_\alpha^{(\Delta)}(\bm
y, t; A)$ for $\alpha = 1, 2$ is given by
\begin{eqnarray}
  \lefteqn{
    i \partial_t
    \left\{
    \hat\phi_0^{(\Delta)}(\bm x,t; A)
    \hat\phi_\alpha^{(\Delta)}(\bm y, t; A),
    \right\}
  }
  \label{eq:time-derivative-phiphi}
  \\\nonumber
  &=&
  \hat U_\Delta(0, t; A)
  \left[
    \hat\phi_0(\bm x)
    \hat\phi_\alpha(\bm y),
    \hat H_\Delta[A_t]
    \right]
  \hat U_\Delta(t, 0; A).
\end{eqnarray}
The commutator on the r.h.s. is arranged as
\begin{eqnarray}
  \lefteqn{
    \left[
      \hat\phi_0(\bm x)
      \hat\phi_\alpha(\bm y),
      \hat {\mathbb P}_\Delta
      \hat H[A_t]
      \hat {\mathbb P}_\Delta
      \right]
  }
  \\\nonumber
  &=&
  \hat {\mathbb P}_\Delta
  \hat H[A_t]
  \left[
    \hat\phi_0(\bm x)
      \hat\phi_\alpha(\bm y),
    \hat {\mathbb P}_\Delta
    \right]
  \\\nonumber
  &+&
  \hat {\mathbb P}_\Delta
  \left[
    \hat\phi_0(\bm x)
    \hat\phi_\alpha(\bm y),
    \hat H[A_t]
    \right]
  \hat {\mathbb P}_\Delta
  +
  \left[
    \hat\phi_0(\bm x)
    \hat\phi_\alpha(\bm y),
    \hat {\mathbb P}_\Delta
    \right]
  \hat H[A_t]
  \hat {\mathbb P}_\Delta.
\end{eqnarray}
By noting $\langle 0|\hat U_\Delta(0, t; A) = \langle 0|$ and $\langle
0| \hat H[A_t]  = 0$, we calculate  a matrix element of  both sides of
\Eq{eq:time-derivative-phiphi}  between  $\langle  0|$  and  $|n,  \bm
P\rangle \in {\mathcal F}_\Delta$ to have
\begin{eqnarray}
  i\partial_t
  \psi_\alpha^{(\Delta)}(\bm x, \bm y, t; A|n, \bm P)
  &=&
  \left\langle 0\left|
  \left[
    \hat \phi_0(\bm x)
    \hat \phi_\alpha(\bm y),
    \hat H[A_t]
    \right]
  \hat U_\Delta(t, 0; A)
  \right| n, \bm P\right\rangle
  \label{eq:B3}
  \\\nonumber
  &+&
  \left\langle 0\left|
  \left[
    \hat \phi_0(\bm x)
    \hat \phi_\alpha(\bm y),
    \hat {\mathbb P}_\Delta
    \right]
  \hat H[A_t]
  \hat {\mathbb P}_\Delta
  \hat U_\Delta(t, 0; A)
  \right| n, \bm P\right\rangle.
\end{eqnarray}
The first term on the r.h.s. reduces to
\begin{eqnarray}
  \mbox{The 1st term}
  &=&
  \left(
  - \frac{\bm \partial_{\bm x}^2}{2m}
  - \frac{\bm D_{\bm y}^2}{2m}
  - A_0(\bm y,t)
  + \Delta \delta_{\alpha 2}
  \right)
  \psi_\alpha^{(\Delta)}(\bm x,\bm y, t; A|n, \bm P)
  \label{eq:B4}
  \\\nonumber
  &+&
  V_{\alpha 1}(\bm x - \bm y)
  \psi_1^{(\Delta)}(\bm x,\bm y, t; A|n, \bm P)
  +
  V_{\alpha 2}(\bm x - \bm y)
  \psi_2^{(\Delta)}(\bm x,\bm y, t; A|n, \bm P),
\end{eqnarray}
while the second term on the r.h.s. is arranged as
\begin{eqnarray}
  \mbox{The 2nd term}
  &=&
  -
  \left\langle 0\left|
  \hat \phi_0(\bm x)
  \hat \phi_\alpha(\bm y)
  \left(
  \hat {\mathbb I}
  -
  \hat {\mathbb P}_\Delta
  \right)
  \hat H[A_t]
  \hat {\mathbb P}_\Delta
  \hat U_\Delta(t, 0; A)
  \right| n, \bm P\right\rangle
  \label{eq:B5}
  \\\nonumber
  &=&
  -
  \int d^3x'\,d^3y'\,
  \sum_{\beta = 1,2}
  \left\langle 0\left|
  \hat \phi_0(\bm x)
  \hat \phi_\alpha(\bm y)
  \left(
  \hat {\mathbb I}
  -
  \hat {\mathbb P}_\Delta
  \right)
  \hat H[A_t]
  \hat {\mathbb P}_\Delta
  \hat \phi_0^\dagger(\bm x')
  \hat \phi_\beta^\dagger(\bm y')
  \right| 0 \right\rangle
  \\\nonumber
  &\times&
  \psi_\beta^{(\Delta)}(\bm x',\bm y', t; A| n, \bm P),
\end{eqnarray}
where the last  line is obtained by using the  following expression of
the identity operator in the subspace $\mathcal F$ as
\begin{equation}
  \hat{\mathbb I}_{\mathcal F}
  =
  \sum_{\beta=1,2}
  \int d^3x'\,d^3y'\,
  \hat\phi_0^\dagger(\bm x')
  \hat\phi_\beta^\dagger(\bm y')
  |0\rangle
  \langle 0|
  \hat\phi_0(\bm x')
  \hat\phi_\beta(\bm y').
\end{equation}
By inserting \Eq{eq:B4}  and \Eq{eq:B5} into \Eq{eq:B3},  we arrive at
\Eq{eq:projected-schroedinger}.

\section{Proof that $\mathcal V(\bm x, \bm y; \bm x',\bm y'; A_t)$ satisfies Schr\"odinger eq. in the external field}
\label{sec:proof_mathcal_v}

We give a proof of \Eq{HALQCD-eq}, i.e., the Schr\"odinger equation in
the external  field with the potential  $\mathcal V(\bm x, \bm  y; \bm
x',\bm  y';  A_t)$  of  \Eq{eq:halqcd_potential_in_external_field}  is
satisfied by the truncated NBS wave functions.
For this purpose,  we first note that the truncated  NBS wave function
is rewritten as
\begin{eqnarray}
  \lefteqn{
    \psi_\alpha^{(\Delta)}(\bm x,\bm y, t; A| n,\bm P)
  }
  \label{eq:truncated_nbs_factorization}
  \\\nonumber
  &=&
  \sum_{n'}^{\widetilde E_{n'}<\Delta}
  \int \frac{d^3 P'}{(2\pi)^3}
  \left\langle 0 \left|
  \hat\phi_0(\bm x)
  \hat\phi_\alpha(\bm y)
  \right| n',\bm P'\right\rangle
  \left\langle
  n', \bm P'
  \left|
  \hat U_\Delta(t, 0; A)
  \right|
  n, \bm P
  \right\rangle
  \\\nonumber
  &=&
  \sum_{n'}^{\widetilde E_{n'}<\Delta}
  \int \frac{d^3 P'}{(2\pi)^3}
  \widetilde \psi_\alpha(\bm x-\bm y|n')
  \exp\left( i\bm P'\cdot \frac{\bm x + \bm y}{2} \right)
  \left\langle
  n', \bm P'
  \left|
  \hat U_\Delta(t, 0; A)
  \right|
  n, \bm P
  \right\rangle.
\end{eqnarray}
\Eq{HALQCD-eq}    reduces   to    the    coupled   channel    equation
\Eq{eq:projected-schroedinger} in the following way:
\begin{eqnarray}
  \lefteqn{
    \mbox{r.h.s. of \Eq{HALQCD-eq}}
  }
  \\\nonumber
  &=&
  \int d^3r'\, d^3R'\,
  \int d^3 x''\,d^3 y''\,
  \sum_{m}^{\widetilde E_m < \Delta}
  \\\nonumber
  &\times&
  \left\{
  V_{11}(\bm x, \bm y; \bm x'',\bm y''; A_t)
  \widetilde \psi_1(\bm x'' - \bm y''| m)
  +
  V_{12}(\bm x, \bm y; \bm x'',\bm y''; A_t)
  \widetilde \psi_2(\bm x'' - \bm y''| m)
  \right\}
  \\\nonumber
  &\times&
  \widetilde \psi_1^\vee(\bm r'| m)
  \delta^3\left(
  \frac{\bm x'' + \bm y''}{2}
  -
  \bm R'
  \right)
  \\\nonumber
  &\times&
  \sum_{n'}^{\widetilde E_{n'} < \Delta}
  \int \frac{d^3 P'}{(2\pi)^3}
  \widetilde \psi_1(\bm r'|n')
  \exp\left(i \bm P'\cdot \bm R'\right)
  \left\langle n',\bm P'\left|
  \hat U_\Delta(t, 0; A)
  \right| n, \bm P\right\rangle
  \\\nonumber
  &=&
  \int d^3 x''\,d^3 y''\,
  \sum_{m}^{\widetilde E_m < \Delta}
  \\\nonumber
  &\times&
  \left\{
  V_{11}(\bm x, \bm y; \bm x'',\bm y''; A_t)
  \widetilde \psi_1(\bm x'' - \bm y''| m)
  +
  V_{12}(\bm x, \bm y; \bm x'',\bm y''; A_t)
  \widetilde \psi_2(\bm x'' - \bm y''| m)
  \right\}
  \\\nonumber
  &\times&
  \int \frac{d^3 P'}{(2\pi)^3}
  \exp\left(i \bm P'\cdot \frac{\bm x'' + \bm y''}{2} \right)
  \left\langle m,\bm P'\left|
  \hat U_\Delta(t, 0; A)
  \right| n, \bm P\right\rangle
  \\\nonumber
  &=&
  \int d^3 x''\,d^3 y''\,
  \begin{array}[t]{l}
    \left\{
    V_{11}(\bm x, \bm y; \bm x'',\bm y''; A_t)
    \widetilde \psi_1^{(\Delta)}(\bm x'', \bm y'', t; A| n, \bm P)
    \right.
    \\[2ex]
    + \left.
    V_{12}(\bm x, \bm y; \bm x'',\bm y''; A_t)
    \widetilde \psi_2^{(\Delta)}(\bm x'', \bm y'', t; A| n, \bm P)
    \right\},
  \end{array}
\end{eqnarray}
where,     to     obtain     the      second     line,     we     used
\Eq{eq:truncated_nbs_factorization} and introduced  $\bm r' \equiv \bm
x' - \bm y'$  and $\bm R' \equiv (\bm x' + \bm  y')/2$.
To obtain the third line, we completed the integration of $\bm r'$ and
$\bm R'$  by using  the orthogonality  relation of  the dual  basis in
\Eq{eq:orthogonality}.      To      obtain     the      last     line,
\Eq{eq:truncated_nbs_factorization} was used again.

\section{The functional derivative of the evolution operator by the external field}
\label{sec:evolution_operator}

We derive the formula:
\begin{equation}
  \left.
  \frac{\delta \hat U_\Delta(t, 0; A)}{\delta A_{\mu}(z)}
  \right|_{A=0}
  =
  i e^{(t-z_0)\hat H}
  \hat{\mathbb P}_\Delta
  \hat j^{\mu}_{p}(\bm z)
  \hat{\mathbb P}_\Delta
  e^{iz_0 \hat H}.
  \label{eq:der_U_der_a}
\end{equation}
From \Eq{eq:projected-time-evolution}, we have
\begin{eqnarray}
  \lefteqn{
    \frac{\delta \hat U_\Delta(t,0; A)}{\delta A_{\mu}(z)}
  }
  \\\nonumber
  &=&
  \sum_{n=1}^{\infty}
  (-i)^n
  \sum_{j=1}^{n}
  \int_0^t dt_n
  \cdots
  \int_0^{t_2} dt_1\,
  \hat H_\Delta[A_{t_n}]
  \cdots
  \hat H_\Delta[A_{t_{j+1}}]
  \frac{\delta \hat H_\Delta[A_{t_j}]}{\delta A_{\mu}(z)}
  \hat H_\Delta[A_{t_{j-1}}]
  \cdots
  \hat H_\Delta[A_{t_1}]
  \\\nonumber
  &=&
  -i
  \int_0^t dt'\,
  \hat U_\Delta(t,t'; A)
  \frac{\delta \hat H_\Delta[A_{t'}]}{\delta A_{\mu}(z)}
  \hat U_\Delta(t',0; A).
\end{eqnarray}
By using
\begin{equation}
  \left.\frac{\delta \hat H[A_{t'}]}{\delta A_{\mu}(z)}\right|_{A= 0}
  =
  -\hat j^{\mu}_{\rm p}(\bm z)\delta(t' - z_0),
  \label{eq:der_h_der_a}
\end{equation}
and
\begin{equation}
  U_\Delta(t,s; A=0)
  \hat{\mathbb P}_{\Delta}
  =
  e^{i(t-s)\hat H} \hat{\mathbb P}_\Delta,
\end{equation}
we are left with \Eq{eq:der_U_der_a}.

\section{Green's function of the effective quantum mechanics}
\label{sec:green_function}
The retarded  Green's function of the  effective Hamiltonian $\mathcal
H$ is defined as the solution to the differential equation:
\begin{equation}
  \left(
  i\frac{\partial}{\partial t}
  - {\mathcal H}
  \right)
  G(\bm x,\bm y, t; \bm x',\bm y', t')
  =
  \delta(t - t')
  \delta^3(\bm x - \bm x')
  \delta^3(\bm y - \bm y'),
\end{equation}
with the boundary condition
\begin{equation}
  \lim_{t\to -\infty}
  G(\bm x,\bm y, t; \bm x',\bm y', t')
  =
  0.
\end{equation}
The solution is explicitly given as
\begin{eqnarray}
  \lefteqn{
    G(\bm x,\bm y,t; \bm x', \bm y', t')
  }
  \label{eq:green_function}
  \\\nonumber
  &\equiv&
  -i\theta(t- t')
  \sum_{n=0}^\infty
  \int \frac{d^3P}{(2\pi)^3}
  \chi^{\rm R}_{n,\bm P}(\bm x,\bm y)
  \chi^{\rm L}_{n,\bm P}(\bm x',\bm y')
  e^{-i{\mathcal E}_n(\bm P^2)(t - t')},
\end{eqnarray}
where $\chi^{\rm L}_{n,\bm P}(\bm x,\bm  y)$ and $\chi^{\rm R}_{n, \bm
  P}(\bm  , \bm  y)$ are  the left  and the  right eigen  functions of
$\mathcal H$ associated with the eigenvalue ${\mathcal E}_n(\bm P^2)$
given in \Eq{eq:eigenvalue_properties}.
The  retarded  Green's  function  is used  to  solve  an  differential
equation
\begin{equation}
  \left( i\partial_t - {\mathcal H} \right)
  F(\vec x,\vec y, t)
  =
  f(\vec x,\vec y, t),
\end{equation}
with a boundary condition $\lim_{t\to-\infty} F(\vec x,\vec y, t) = 0$.
Its solution is given as
\begin{equation}
  F(\vec x,\vec y,t)
  =
  \int dt'
  \int d^3x'\,d^3y'\,
  G(\bm x,\bm y, t; \bm x',\bm y', t')
  f(\bm x',\bm y',t').
\end{equation}
\end{document}